\documentclass[12pt,showpacs,preprintnumbers,amsmath,amssymb]{revtex4}

\usepackage{amsmath} \usepackage{graphicx} \usepackage{amsfonts}
\usepackage{array} \usepackage{amsthm}

\usepackage[dvips]{epsfig}
\usepackage{latexsym}
\usepackage{psfrag}

\newcommand{\be}{\begin{equation}}
\newcommand{\ee}{\end{equation}}
\newcommand{\bea}{\begin{eqnarray}}
\newcommand{\eea}{\end{eqnarray}}

\newcommand{\vf}{\varphi} 
 
\newcommand{\mn}{{\mu\nu}}
\newcommand{\Lef}{\left(} \newcommand{\Rig}{\right)}
\newcommand{\p}{\partial} \newcommand{\ep}{\varepsilon}
\newcommand{\bR}{\mathbb{R}}

\begin{document}
\title{Einstein--Yang--Mills strings}
\author{
  Dmitri V. Gal'tsov}
\affiliation{Department of Theoretical
Physics, Moscow State University, 119899, Moscow, RUSSIA}
\author{
    Evgeny A. Davydov}
\affiliation{Department of Theoretical
Physics, Moscow State University, 119899, Moscow, RUSSIA}
\author{ Mikhail S. Volkov}\affiliation{ Laboratoire de Math\'ematiques et Physique Th\'eorique,
CNRS-UMR 6083, Universit\'e de Tours, Parc de Grandmont, 37200
Tours, FRANCE} 

\begin{abstract}
We present globally regular vortex-type solutions for a pure SU(2)
Yang-Mills field  coupled to gravity  in 3+1 dimensions. These
gravitating vortices are  static, cylindrically symmetric and
purely magnetic, and they support a non-zero chromo-magnetic flux
through their cross section. In addition, they carry a constant
non-Abelian current, and so in some sense they are analogs of the
superconducting cosmic strings. They have a compact central core
dominated by a longitudinal magnetic field and  endowed with
an approximately Melvin geometry. This magnetic field component
gets color screened in the exterior part of the core, outside of
which the fields approach exponentially fast those of the
electrovacuum Bonnor solutions with a circular magnetic field. In
the far field zone the solutions are not asymptotically flat but
tend to vacuum Kasner metrics.
\end{abstract}

\maketitle

Classical solutions in non-Abelian gauge models coupled to gravity
in four space-time dimensions   attracted a lot of attention
 after the discovery by Bartnik and McKinnon
\cite{Bartnik:1988am} of particle-like solutions  in the
Einstein-Yang-Mills (EYM) theory and after the  construction of the
corresponding black holes (see
\cite{Volkov:1998cc} for a review). Since then a variety of static
and  stationary axisymmetric EYM solutions have been considered,
but static, cylindrically symmetric EYM configurations have remained
unexplored for some reason.
To fill this gap, we consider in this letter static, cylindrically symmetric,
purely magnetic, globally regular four-dimensional EYM solutions
of cosmic string type.

Surprisingly, we find strings endowed with a constant current
along their symmetry axis, and so in some sense they can be
considered as analogs of Witten's superconducting cosmic strings
\cite{Witten}. In the latter case the superconductivity arises due
to the presence of two complex scalars in the theory. In our case
their role is played by two color components of the
self-interacting Yang-Mills field, which behave exactly in the same
way as the scalars in Witten's model. Specifically, one of them
develops a non-zero condensate value in the vortex core and
vanishes  at infinity, while  the other  one behaves the other
way round. Since the gauge symmetry of our theory is not broken,
the gauge field coupled to the current is long-range. The slow
fall-off of the energy density  in the direction orthogonal to
the string then does not let the solutions to be asymptotically
flat, although they are asymptotically Ricci flat. This is the
principal difference of the EYM strings  as compared to the
previously studied  currentless Abelian self-gravitating cosmic
strings of the Nielsen-Olesen type \cite{Garfinkle}.

In what follows we present a numerical evidence for the existence
of a family of globally regular EYM string solutions labeled by an
integer winding number $\nu$ and by a real parameter $p$
determining the value of the magnetic field at the symmetry axis.
These EYM strings are filled with a self-interacting
chromo-magnetic field giving rise to a constant current along the
string symmetry axis and also to a chromo-magnetic flux through the
string cross section. The flux and current can be defined in a
gauge-invariant way within the approach of asymptotic symmetries
\cite{AD}. The geometry of the solutions interpolates between flat
Minkowskian geometry at the symmetry axis and vacuum Kasner
geometry in the far field zone. Interestingly, these solutions
survive even if the coupling to gravity is off and the metric is
flat everywhere. They describe then `pure Yang-Mills
superconducting strings' with a regular central core outside of
which there remains   only a U(1) component of the gauge field,
which, being coupled to the current, diverges logarithmically for
large $r$. As a result, the total energy per unit length for the
flat space solutions diverges at large $r$. Getting back to curved
space, this divergence is cured by gravity.

We consider the SU(2) Einstein-Yang-Mills (EYM) theory
\be            \label{0}
S=\int d^4 x \sqrt{-g}\Lef
-\frac{\bR}{16\pi G}-\frac{1}{4e^2}\,F_{\mn}^a F^{a\mn}\Rig,
\ee
where
$F_{\mn}^a=\p_\mu A_{\nu}^a-\p_\nu
    A_{\mu}^a+\ep_{abc}A_{\mu}^b A_{\nu}^c$
and $e$ is the gauge coupling constant. With $A=\frac12\tau_a A^a_\mu dx^\mu$
where $\tau_a$ are the Pauli matrices, the
local SU(2) gauge symmetry of the action is expressed by
 $A\to U(A+id)U^{-1}$ where
$U(x)\in SU(2)$. We shall be interested in static, cylindrically
symmetric systems invariant under the action of the three
commuting Killing vectors $\p_t,\p_\vf,\p_z $. The spacetime
metric then can be parameterized as \be
\label{1}
ds^{2}=l^2\{N(r)^2dt^2-H(r)^2dr^2-L(r)^2d\vf^2-K(r)^2\,dz^2\}, \ee
where $l$ is the length scale and all the other quantities are
dimensionless. For the gauge field  we make the following purely
magnetic ansatz, \be                             \label{2}
A=\tau_2\, R(r)\, dz+\tau_3\,P(r)\,d\vf. \ee Using
Eqs.\eqref{1},\eqref{2} we obtain $S=\frac{1}{\kappa e^2}\int d^4x
{\cal L}$ where $\kappa={8\pi G}/(el)^2$ and \be \frac{H{\cal
L}}{LNK}=\frac{N'K'}{NK}+\frac{N'L'}{NL}+\frac{L'K'}{LK}
-\frac{\kappa R^{\prime 2}}{2K^2} -\frac{\kappa P^{\prime
2}}{2L^2} -\frac{\kappa H^2R^2 P^2}{2K^2L^2}. \ee We notice that
the reduced Lagrangian ${\cal L}$ is invariant under $P\to\alpha
P$, $L\to\alpha L$, $R\to\beta R$, $K\to\beta K$, $N\to N$, $H\to
H$, with constant $\alpha,\beta$. It also admits  a discrete
symmetry $P\leftrightarrow R$, $L\leftrightarrow K$,  which is
equivalent to $z\leftrightarrow \varphi$. Varying ${\cal L}$ gives
the  field equations, which can be represented in the form \bea
\label{3} N''&=&\left(\frac{N^{\prime
2}}{N^2}+\frac{L'S'}{LS}-\frac{L^{\prime 2}}{L^2}
+\kappa\,\frac{P^2R^2}{S^2}\right)N\,,\nonumber \\
L''&=&\left(\frac{N'S'}{NS}-\kappa\,\frac{P^{\prime 2}}{L^2}\right)L\,,\nonumber\\
S''&=&-\kappa\,\frac{N^2 R^2 P^2}{S}\,,\nonumber\\
\left(\frac{L^2}{S}\,R'\right)'&=&\frac{N^2P^2}{S}\,R\,,\nonumber\\
\left(\frac{S}{L^2}\,P'\right)'&=&\frac{N^2R^2}{S}\,P\,, \eea plus
a first order constraint equation \be \label{4} \frac{N'S'}{N
S}+\frac{L'S'}{L S}-\frac{L^{\prime 2}}{L^2}
=\frac{\kappa}{2}\left(
 \frac{P'^2}{L^2}+\frac{L^2 R'^2}{S^2}
    -\frac{N^2 P^{2}R^{2}}{S^2}\right).
\ee
Here $S=LK$ and, after varying, we have imposed the coordinate condition $H=N$.
The energy density is
\be                                                 \label{T00}
T^0_0=\frac{P^{\prime 2}}{2N^2L^2}+\frac{R^{\prime 2}}{2N^2K^2}+\frac{P^2R^2}{2S^2},
\ee
while the other non-trivial components of $T^\mu_\nu$ are obtained by choosing
the signs in front of the three terms in this formula as
$T^r_r\sim (-,-,+)$,  $T^z_z\sim (+,-,-)$, $T^\varphi_\varphi\sim (-,+,-)$,
such that $T^\mu_\mu=0$.
Associated with the two continuous global symmetries of ${\cal L}$
there are two Noether charges
\begin{eqnarray}                      \label{5}
Q_1 &=& S\Lef \frac{N'}{N}-\frac{L'}{L}- \kappa\,\frac{PP'}{L^2}\Rig \nonumber,\\
Q_2 &=& S\Lef \frac{N'}{N}-\frac{S'}{S}+\frac{L'}{L}-
  \kappa\,\frac{L^2RR'}{S^2}\Rig,
\end{eqnarray}
whose conservation can be checked straightforwardly.

We want the fields to be regular at the symmetry axis $r=0$. This
implies that one should have $P(0)=\nu\in \mathbb{Z}$, since in
this case, with $U=e^{-\frac{i\nu}{2}\tau^3\varphi}$, one can pass
to the gauge where the gauge field will be regular at the axis:
$2A=U\tau^2 U^{-1}\,R\, dz+\tau_3\,(P-\nu)\,d\vf$. The metric
should become  Minkowskian for $r\to 0$.  The most general local
power series solution of Eqs.\eqref{3} with such boundary
conditions reads \bea                         \label{6}
P&=&\nu-pr^2+O(r^4),~~~R=qr^\nu+O(r^{\nu+2}),~~~ \\
N&=&1+O(r^2),~~~S=r+O(r^3),~~~
L=r+O(r^3),  \nonumber
\eea
where $p,q$ are two integration constants.
This solution fulfills also the constraint
\eqref{4}.
Inserting \eqref{6} to \eqref{5} fixes the values of the Noether charges:
$Q_1=2\nu\kappa p-1$ and $Q_2=0$.   We can now integrate the
equations starting from the origin on, with the boundary condition \eqref{6},
and this reveals the following picture. Fixing a value of $p$ and
varying $q$, the numerical solution generically terminates at a finite
point $r=r_\ast(q)$ where $P'$ becomes large and either positive or negative,
depending on $q$. Adjusting properly the value of $q$, we can postpone
further and further
 the moment $r_\ast(q)$ where the numerical procedure
crashes.
We thus extend the solution from the axis farther and farther to the
asymptotic region, and we observe then that the amplitude $P$ approaches
zero very quickly with growing $r$. As a result, we conclude that
at large $r$
the globally regular solutions approach configurations with $P=0$.

\begin{figure}[t]
\hbox to\linewidth{\hss%
  \psfrag{x}{$\xi=\ln(1+r)$}
  \psfrag{L}{$L$}
 \psfrag{N}{$N$}
\psfrag{P}{$P$}
\psfrag{R}{$\frac12\,R$}
\psfrag{K}{$K$}
\psfrag{T00}{$4T^0_0$}
    \resizebox{9cm}{5cm}{\includegraphics{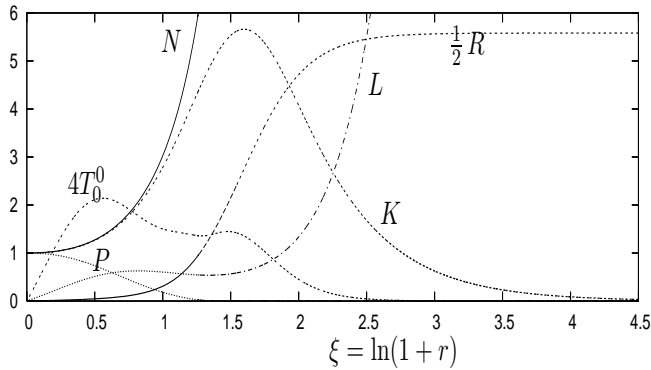}}%
\hss} \caption{\label{fig1}\small The globally regular solution of
Eqs.\eqref{3} with $p=0.8$, $\nu=1$. The central core extends up
to $\xi\approx 1.2$, after which the configuration practically
coincides with the Abelian solution \eqref{7} of Bonnor with
$m=-1.94$. For $\xi\geq 3$ the gauge field dies out and the
solution becomes Kasner. }
\end{figure}

Setting $P=0$ in Eqs.\eqref{3}, the SU(2) gauge field configuration becomes U(1),
in which case the general solution of Eqs.\eqref{3},
let us call it $N_0,L_0,S_0,R_0$, reads \cite{B}
\bea                              \label{7}
L_0&=&a_1 x^{1+m}+a_2 x^{1-m},~~~N_0=a_3 x^{m^2-1}L_0, \\
S_0&=&a_4\, x,~~~R_0=a_5+a_6\,\frac{x^{1-m}}{L_0}\,,~~~x=r-r_0\,,\nonumber
\eea
where $m,a_1,\ldots a_6,r_0$ are 8 integration constants.
Imposing the constraint \eqref{4} would imply also that
$a_6\sqrt{\kappa a_1}=\pm a_4\sqrt{2a_2}$, but we do not demand this at the moment.
We conclude that at large $r$ our solution is given by
\bea                       \label{8}
L&=&L_0+\delta L,~~~N=N_0+\delta N,~~~S=S_0+\delta S,\nonumber \\
R&=&R_0+\delta R,~~~P=\delta P, \eea where the deviations $\delta
L,\ldots ,\delta P$ vanish as $r\to\infty$. To determine them we
insert \eqref{8} to Eqs.\eqref{3} and linearize with respect to
the deviations. This gives \be                        \label{9}
\delta P=a_7\frac{L_0}{\sqrt{S_0}} \exp\left(-\frac{L_0 N_0
R_0}{S_0}\,x\right)(1+\ldots), \ee where $a_7$ is a new
integration constant and the dots stand for the subleading terms.
It follows also that $\delta N$, $\delta L$, $\delta S$, and $\delta R$
are all of the order $(\delta P)^2$ and no new integration
constants appear in these variations, since the background
solution $N_0,L_0,S_0,R_0$ already contains the maximal number of
integration constants.

\begin{figure}[t]
\hbox to\linewidth{\hss%
  \psfrag{x}{$p$}
  \psfrag{m1}{$m(1,p)$}
 \psfrag{C1}{${\cal I}(1,p)$}
\psfrag{MASS1}{$M(1,p)$}
  \psfrag{m2}{$m(2,p)$}
 \psfrag{C2}{${\cal I}(2,p)$}
\psfrag{MASS2}{$M(2,p)$}
    \resizebox{9cm}{5cm}{\includegraphics{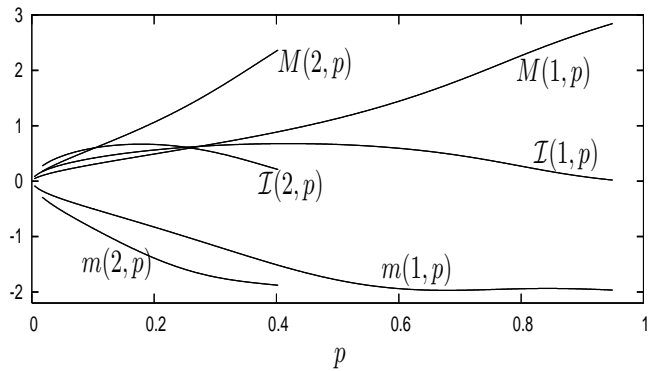}}%
\hss}
\caption{\label{fig2}\small The parameter $m(\nu,p)$, the Tolman mass
$M(\nu,p)$ and the current ${\cal I}(\nu,p)\equiv I_2/(2\pi)$ for the
globally regular solutions with $\nu=1,2$.
}
\end{figure}

Summarizing, the local solution at large $r$ contains 9 free parameters
 and is given by Eqs.\eqref{7}--\eqref{9},
while that at small $r$ contains 2 integration constants in
 Eq.\eqref{6}. We then integrate numerically Eqs.\eqref{3} to
extend these local solutions
to the intermediate region where
we match them. This gives us global solutions in the interval
$r\in [0,\infty)$. To fulfill the 10 matching conditions for the 5
functions in Eqs.\eqref{3} and for their first derivatives, we have
in our disposal 2+9=11 free parameters,
and so there is one free parameter left after the matching.
This parameter,  we choose it to be
$p$ in  Eq.\eqref{6}, labels the global solutions.

 No special care is to be taken about the
constraint \eqref{4}, since being imposed at $r=0$ it `propagates' to
all values of $r$. As a result, the values of the
asymptotic parameters of the global solutions found by the
matching
will automatically satisfy
$a_6\sqrt{\kappa a_1}=\pm a_4\sqrt{2a_2}$.
Similarly, the axis values of the Noether charges $Q_1=2\nu\kappa p-1$, $Q_2=0$
propagate to the large $r$  region  and impose the asymptotic conditions
\be                 \label{QQ}
a_4(m^2-1)=Q_1,~~~~
(2+m)a_4+2\sqrt{2\kappa a_1 a_2}\,a_5=0.
\ee

We thus obtain a family of globally regular solutions of Eqs.\eqref{3}
labeled by a real parameter $p>0$ and by an integer $\nu$,
and we use the length scale for which $\kappa=2$.
The typical solution is shown in Fig.1. These EYM strings have a compact
central core filled with a chromo-magnetic field, outside of which
 $P\approx 0$ and the
remaining gauge field becomes effectively Abelian. The interior par of the core,
where $R\approx 0$, is also
effectively Abelian and can be well described by the Melvin solution. The latter
can be obtained from the general solution \eqref{7} by applying the
$z\leftrightarrow\varphi$ symmetry and adjusting the constants such that
the boundary conditions at $r=0$ are fulfilled:
\be
N=K=1+\frac{\kappa}{2}\,p^2r^2,~~L=\frac{r}{N},~~P=\nu-\frac{pr^2}{N}.
\ee
Since
this inner solution has a magnetic field along the
$z$ axis, we call it z-string. The metric in this limit acquires
an additional symmetry with respect to Lorentz boosts in the $(0,z)$ plane,
and the energy-momentum tensor in this
limit is such that $T_0^0=T_z^z=-T^r_r=-T^\varphi_\varphi$.

The outer part of the central core is the transition region where
the non-Abelian effects become essential, since the $R$ field
starts growing there, playing a role of the effective mass term
for the $P$ field, and so the latter starts falling down to zero
exponentially fast. As a result, outside of the central core one
has $P\to 0$ and the whole configuration approaches exponentially
fast one of the Abelian Bonnor solutions \eqref{7}. These, in
their turn, contain an intermediate region filled with a circular
magnetic field, which we call $\phi$-string and which has
$T_0^0=T_\varphi^\varphi=-T^r_r=-T^z_z$. This magnetic field falls
off polynomially for large $r$, and in the far field zone the
solutions  approach asymptotically  vacuum Kasner metrics.
Summarizing, the whole solution can be viewed as a non-linear
superposition of a z-string placed at small $r$ and a
$\phi$-string located at larger $r$, which is demonstrated by the
profile of the radial energy density, $SN^2T^0_0$, clearly showing
two peaks (see Fig.1).
\begin{figure}[t]
\hbox to\linewidth{\hss%
  \psfrag{x}{$X$}
  \psfrag{z}{$Z$}
    \resizebox{9cm}{5cm}{\includegraphics{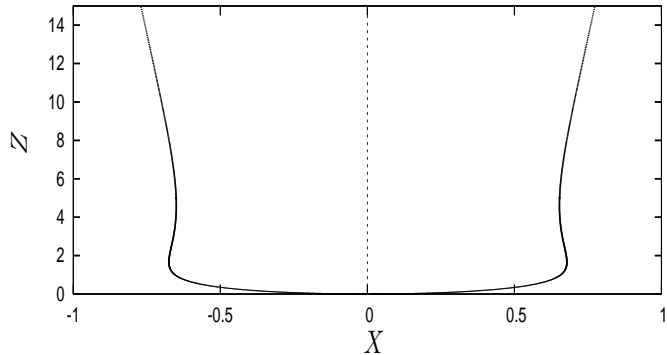}}%
\hss}
\caption{\label{fig3}\small The Euclidean 2-geometry of the $(r,\varphi)$
string orthogonal section can be realized on a 2-surface of revolution
in three dimensional Euclidean space obtained by rotating
the shown planar curve
around the Z-axis.
The curve shown corresponds to the string
with $\nu=1,p=0.75$.
}
\end{figure}

It is natural to wonder if the solutions can be asymptotically flat,
as in the case of gravitating Nielsen-Olesen strings \cite{Garfinkle}.
However,
taking a linear combination of the first equation in \eqref{3}
with the constraint \eqref{4} and integrating gives
\be                       \label{10}
\lim_{r\to\infty}S\,\frac{N'}{2N} =\int_0^\infty SN^2
T^0_0dr.
\ee
The integral on the right here is manifestly positive and equals
the Tolman mass, $M$, while the
left hand side
is equal to $(m^2+|m|)a_4$.
As a result, one always has $m\neq 0$ and so solutions with $T_0^0\neq 0$ cannot
approach flat metric for large $r$.

Apart from the mass $M$, another important parameter of the EYM
strings is the current flowing along them. Although there is no
Higgs field in the theory, it is still possible to introduce
conserved and gauge invariant currents within the asymptotic
symmetries approach of Abbott and Deser \cite{AD}. Specifically,
since our solutions are asymptotically vacuum, one has
$A=A^\infty+a$, where $A^\infty=iU_\infty dU^{-1}_{\infty}$ and
$a\to 0$ as $r\to\infty$, but otherwise $a$ need not to be small.
Defining ${\cal D}^\infty_\mu a_\nu=\nabla_\mu
a_\nu-i[A^\infty_\mu,a_\nu]$ one can introduce \be \label{f}
f^{(\infty)a}_{\mu\nu}={\text tr} \left[\left({\cal D}^\infty_\mu
a_\nu -{\cal D}^\infty_\nu a_\mu\right)\,U_\infty\tau^a
U_\infty^{-1} \right] \ee and then the conserved and gauge invariant
currents are $J^a_\mu=\nabla^\sigma f^{(\infty)a}_{\sigma\mu}$.
The
total current through the string cross section is then given by
\be
\frac{I_a}{2\pi}=\int_0^\infty J^z_a \sqrt{-g}dr=
\delta_a^2\lim_{r\to\infty}\frac{L^2}{S}R'=
-2m\sqrt{\frac{2}{\kappa}a_1 a_2}\,\delta_a^2.
\ee
Our strings thus carry a persistent current, and for this reason
they can be called
`superconducting'.

At the axis we have similarly $A=A^0+a$, where
$A^0=iU_0 dU^{-1}_{0}$ and $a\to 0$ for $r\to 0$. Defining then
$f^{(0)a}_{\mu\nu}$ as in \eqref{f} but with $`\infty$' replaced by $`0$',
we can calculate its flux through the sting cross section, $\Sigma$,
\be
\int_{\Sigma} f^{(0)a}_{\mu\nu}\, dx^\mu\wedge dx^\nu=-\nu\,\delta^a_3\,,
\ee
and this is `quantized' as for the Nielsen-Olesen vortex.

The mass, current and the parameter $m$ as functions of $p$ are shown
in Fig.2 for solutions with $\nu=1,2$.
We see that $m$ is always negative
and tends to zero for $p\to 0$, in which limit the gauge field becomes pure
gauge with  $R(r)=0$ and $P(r)=\nu$, so its energy vanishes
and the metric becomes flat.
For $p=1/(2\kappa\nu)$, when $Q_1=0$,
Eq.\eqref{QQ} implies that $m=-1$, in which case $g_{00}$ becomes proportional
to $g_{\varphi\varphi}$ for large $r$, and so the metric acquires the asymptotical
boost symmetry in the $(t,\varphi)$ plane.

Another interesting limit is
$\kappa\to 0$ with fixed $p$, in which case the gravity switches off
and the geometry becomes flat. The gauge field, however, remains non-trivial.
Such Minkowski space solutions also
show a regular central core where $R\approx 0$ and $P\neq 0$, outside
of which $P$ is exponentially small, but the large $r$ behavior of $R$
is now different. Specifically, for small $\kappa$ one has
$m= -\sqrt{{\kappa}/{2}}\,{\cal I}$,
$a_1-1/2=O(\sqrt{\kappa})$, $a_2\to 1-a_1$, using which in Eq.\eqref{7}
and taking the limit gives $R=C_0+{\cal I}\ln(r)$ for large $r$.
Of course, the same results are obtained by solving the two Yang-Mills equations
in \eqref{3} with $N=1,L=S=r$.
The amplitude $R$ is thus now logarithmically divergent at infinity, as it should be,
since we effectively have here a U(1) gauge field coupled to the conserved current.
We thus obtain `superconducting strings' made of pure Yang-Mills field in flat space,
and these, to the best of our knowledge, have also never been described
in the literature. Getting back to the self-gravitating case, one can say
that the gravity provides a `dynamical cut-off' by rendering $R$
finite at large $r$, which makes the total energy finite.

It is interesting to visualize the two-dimensional geometry
$dl^2=N^2 dr^2+L^2d\varphi^2$ of the $(r,\varphi)$
plane orthogonal to the string in terms of embeddings.
The same geometry can be realized on a 2-surface of revolution in three
dimensional Euclidean space with coordinates $X,Y,Z$
obtained by rotating around the Z-axis
the planar curve defined parametrically by relations $X=\pm L(r)$, $Z=Z(r)$
with $Z^{\prime 2}=N^2-L^{\prime 2}$.  This defines a `flowerpot' surface
shown in Fig.3. We observe that the circumference of a circle
centered at the
string axis in the plane orthogonal to the string grows up slower than
the radius of the circle.

Summarizing, we have presented superconducting cosmic string
solutions for a pure self-gravitating Yang-Mills field. We have
not studied their stability yet. As a matter of fact, many
solutions in the EYM theory are unstable. However, since we now
have the conserved current, there is a chance that it may
stabilize the solutions.  Other interesting problems to study
would be to consider the black hole generalizations for our
solutions or to include other fields. We also notice that
replacing in all above formulas $t\leftrightarrow z$ gives
solutions with an electric field. They do not have a current, but
possess instead a chromo-electric charge.

The work of D.V.G. and E.A.D. 
was supported in part by the RFBR grant 02-04-16949.

\end{document}